\documentclass{article}
\usepackage{spconf,amssymb,amsmath,epsfig}
\usepackage{textcomp}
\usepackage{amsfonts,amsthm,amsopn}
\usepackage{hyperref}
\usepackage{url}
\usepackage[capitalise]{cleveref}
\usepackage{graphicx,caption,lipsum}
\usepackage{booktabs,multirow}
\usepackage{placeins}
\usepackage{algpseudocode,algorithm}
\usepackage{color}
\usepackage[table,xcdraw]{xcolor}
\usepackage{bbm}

\ninept
\def\reg{{\rm\ooalign{\hfil
      \raise.07ex\hbox{\scriptsize R}\hfil\crcr\mathhexbox20D}}}

\name{Author(s) Name(s)\thanks{Thanks to XYZ agency for funding.}}
\title{Cross-lingual Text-independent Speaker Verification using Unsupervised Adversarial Discriminative Domain Adaptation}

\makeatletter
\def\name#1{\gdef\@name{#1\\}}
\makeatother 
\name{{\em } }

\name{{Wei Xia$^1$, Jing Huang$^2$, John H.L. Hansen$^1$}}

\address{
$^1$ Center for Robust Speech Systems, UT-Dallas, TX, USA \\
$^2$ JD AI Research, Mountain View, CA, USA
{\small \tt}}

\newcommand{\vct}[1]{\boldsymbol{\mathbf{#1}}} 
\newcommand{\mat}[1]{\boldsymbol{\mathbf{#1}}} 

\newcommand{\etal}{et al.}

\graphicspath{{./}}

\begin{document}

\maketitle

\ninept

\begin{abstract}
Speaker verification systems often degrade significantly when there is a language mismatch between training and testing data. Being able to improve cross-lingual speaker verification system using unlabeled data can greatly increase the robustness of the system and reduce human labeling costs. In this study, we introduce an unsupervised Adversarial Discriminative Domain Adaptation (ADDA) method to effectively learn an asymmetric mapping that adapts the target domain encoder to the source domain, where the target domain and source domain are speech data from different languages. ADDA, together with a popular Domain Adversarial Training (DAT) approach, are evaluated on a cross-lingual speaker verification task: the training data is in English from NIST SRE04-08, Mixer 6 and Switchboard, and the test data is in Chinese from AISHELL-I.
We show that with the ADDA adaptation, Equal Error Rate (EER) of the x-vector system decreases from 9.331\% to 7.645\%, relatively 18.07\% reduction of EER, and 6.32\% reduction from DAT as well. Further data analysis of ADDA adapted speaker embedding shows that the learned speaker embeddings can perform well on speaker classification for the target domain data, and are less dependent with respect to the shift in language.

\end{abstract}

\section{Introduction}
\label{sec:intro}


Speaker verification (SV) offers a natural and flexible option for biometric authentication.
The text-independent SV system, which does not require the fixed input voice content, is a flexible and challenging task. 
In real-world scenarios, however, speaker verification systems may degrade significantly when training on one language and test it on another. 
Language mismatch falls into two scenarios that include (i) the speaker verification system is trained on one language, but the enrollment and test data for speakers are in a second language, and (ii) the enrollment data is in one language, but the test data is in a second language. This study focused on the first scenario where the speaker model is trained on English data, but the enrollment and test materials for speakers are in a new language, Chinese. Since it is not desirable to re-train the speaker model on a new language, the challenge is to find an alternative solution which would allow such an existing system to maintain performance when enrollment and test speaker data are from a new language. 

Recently, the speaker representation models have moved from the commonly used i-vector model~\cite{kenny2007joint,matvejka2011full,hansen2015speaker}, with a probabilistic linear discriminant (PLDA) back-end~\cite{kenny2010bayesian,prince2007probabilistic} to a new paradigm: speaker embedding trained from deep neural networks. Various speaker embeddings based on different network architectures~\cite{snyder2018x,michelsanti2017conditional} , attention mechanism~\cite{rahman2018attention,zhang2016end}, loss functions~\cite{wan2018generalized,zhang2018text}, noise robustness~\cite{yu2017adversarial,xia2018speaker}, and training paradigms~\cite{heigold2016end,Heo:2017ci} have been proposed and greatly improve the performance of speaker verification systems. Snyder \etal~\cite{snyder2018x} recently proposed the x-vector model, which is based on a Time-Delay Deep Neural Network (TDNN) architecture that computes speaker embeddings from variable-length acoustic segments. This x-vector model has become very successful in various speaker recognition tasks. We use it as the baseline in this study.

However, models trained with these deep neural networks may not generalize well to other datasets in different domains. To alleviate the domain mismatch problem, we can use domain adaptation methods to reduce the domain shift. We can compensate the mismatch by estimating the compensation model~\cite{aronowitz2014inter,kanagasundaram2015improving,misra2018maximum,misra2018modelling} using unlabeled data and source domain data. 
Adversarial adaptation methods~\cite{ganin2016domain,pei2018multi,yu2017adversarial,chen2016adversarial} were also applied to ensure that the network cannot distinguish the distributions of training and testing examples. Wang \etal~\cite{wang2018unsupervised} proposed an unsupervised approach based on Domain Adversarial Training (DAT) to address speaker recognition problem in domain mismatched conditions.

In this study, we introduce the unsupervised Adversarial Discriminative Domain Adaptation (ADDA)~\cite{tzeng2017adversarial} approach. It was originally tested on image classification tasks.
We adapt the ADDA approach to the cross-lingual unsupervised adaptation for text-independent speaker verification. Unsupervised adaptation without requiring target domain labels largely reduces labeling costs and utilizes a large amount of publicly available online data.
Our approach only requires source and unlabeled target domain data to learn an asymmetric mapping that adapts the target domain feature encoder to the source domain. Furthermore, the ADDA uses separate encoders for the source and target domain without assuming that source and target domain data has a similar class distribution. We show that ADDA is more effective yet considerably simpler than other domain-adversarial methods: the source data is in English from NIST SRE04-08, Mixer 6 and Switchboard, and the target data is in Chinese from AISHELL-I.
We show that with the ADDA adaptation, Equal Error Rate (EER) of the x-vector system decreases from 9.331\% to 7.645\%, relatively 18.07\% reduction on EER. ADDA also has 12.54\% relative reduction of EER compared to DAT.

In the following sections, we describe the ADDA approach and corresponding baseline systems in \cref{sec:model}. We provide detailed explanations of our experiments in \cref{sec:exp}, as well as results and discussions in \cref{sec:result}. Finally we conclude in \cref{sec:conclusion} with future work.

\subsection{Related work}


A number of domain adaptation approaches have been proposed to alleviate the domain shift problem. For example,
Wang \etal~\cite{wang2018unsupervised} apply the DAT technique to alleviate the i-vectors mismatch across different domains. They use a multi-task learning framework to jointly learn a shared feature extractor and two classifiers. With a gradient reversal layer in the domain classifier, the shared feature extractor can extract domain-invariant and speaker-discriminative features. 
  In~\cite{aronowitz2014inter, kanagasundaram2015improving}, the authors proposed an Inter-Dataset Variability Compensation (IDVC) technique to remove the mismatch using Nuisance Attribute Projection (NAP). First, a subspace is computed representing all different data-sets and then NAP is used to remove that subspace as an i-Vector pre-processing step. All these work were on i-vectors for speaker verification, while our work is on the recently proposed x-vectors and shows very promising results.

\begin{figure}[tbp]
    \centering
    \includegraphics[width=1.0\linewidth, height=4cm]{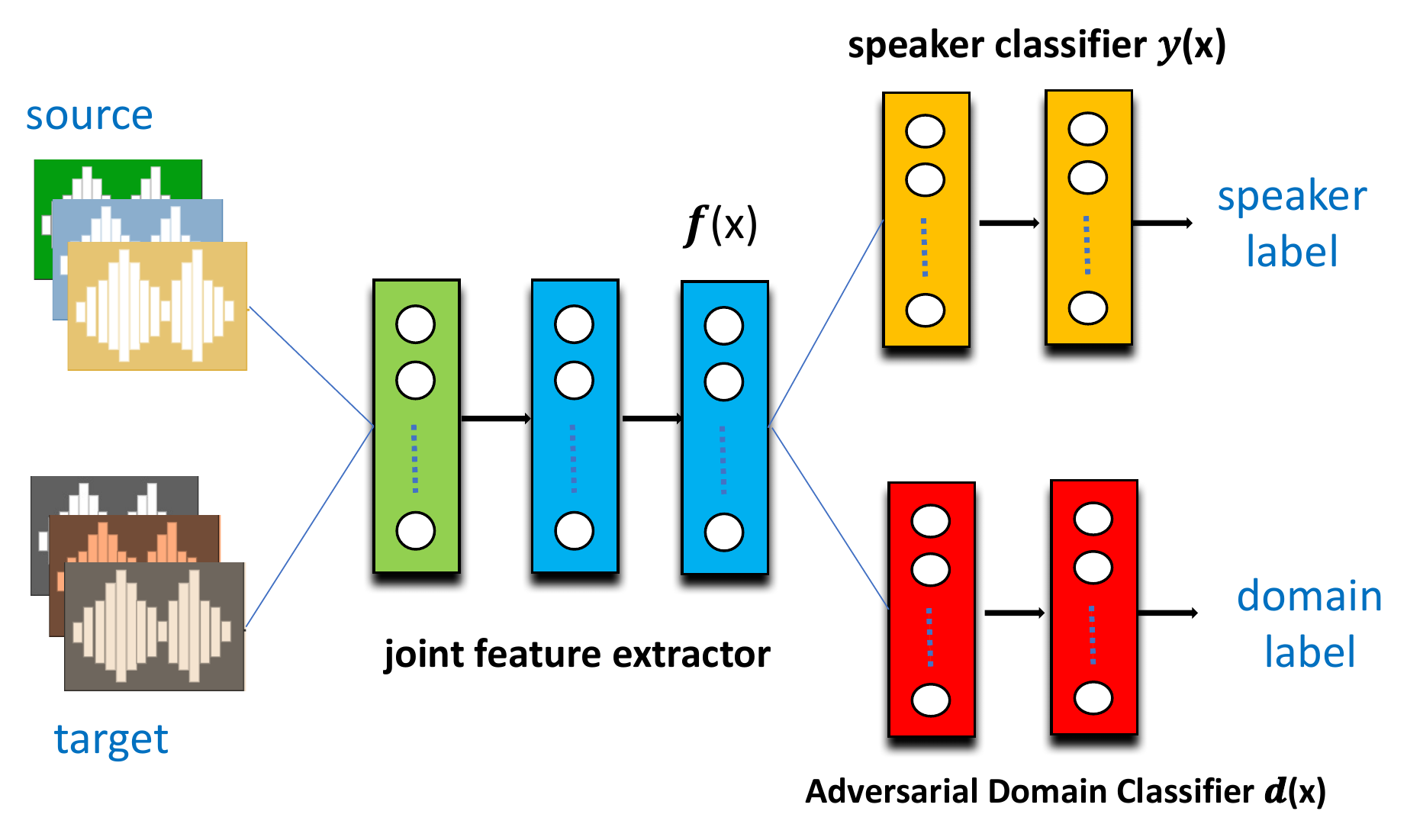}
    \caption{Overview of the Domain Adversarial Training (DAT) framework. Adversarial domain classifier has a gradient reversal layer. Speaker classifier and domain classifier both take input from the joint feature extractor, are optimized to excel in their own tasks.}
    \label{fig:dat}
\end{figure}

\section{Speaker verification systems}
\label{sec:model}

\subsection{The X-vector system}

We use a recently proposed successful speaker model called X-vector~\cite{snyder2018x}, to extract speaker representations, and a Probabilistic Linear Discriminant Analysis (PLDA) back-end to compare pairs of enrollment and test speaker embeddings. The X-vector model is based on a Time-Delay Deep Neural Network (TDNN) architecture that computes speaker embeddings from variable-length acoustic segments. The network consists of layers that operate on speech frames, a statistics pooling layer that aggregates over the frame-level representations, additional layers that operate at the segment-level, and finally a softmax output layer. 
The embeddings are extracted after the statistics pooling layers.


\subsection{Cross-lingual adversarial training baseline}

In order to address the cross-lingual speaker verification problem, we first implement a Domain Adversarial Neural Network (DANN)~\cite{wang2018unsupervised} using Domain Adversarial Training (DAT)~\cite{ganin2016domain} to transfer speaker information from labeled English data to
another language where only unlabeled data exists, for example, Chinese. DANN in~\cref{fig:dat} is a “Y-shaped” network with two discriminative branches: a speaker recognizer and an adversarial language classifier. Both branches take input from a shared feature extractor that aims to learn hidden representations that capture the underlying information of the speaker and are independent of languages. 

We can implement the language independent speaker verification system assuming that DANN can learn features that perform well on speaker classification for the source and target language data, are independent with respect to the shift in language. This can be done by minimizing the speaker classification loss and maximizing the domain classification loss with a gradient reversal layer.
DANN mainly has two components: 1) a speaker recognizer $y$ for the source data; 2) an adversarial language classifier $d$ that predicts a scalar indicating whether the input speech is from the source language or the target language. The two classifiers take input from the shared feature extractor $f$, which operates on the average of the speaker embeddings. The loss function of DANN is a multi-task loss which combines the loss of the speaker classifier and the domain classifier with a weight $\lambda$.
Training DANN consists in optimizing,
\begin{align}
\begin{split}
\mathbb{E}(\theta_f, \theta_y, \theta_d) = 
  \frac{1}{n}\sum_{i=1}^{n}\mathcal{L}_y^{i}(\theta_f, \theta_y) & - \lambda [ \frac{1}{n} \sum_{i=1}^{n}\mathcal{L}_d^{i}(\theta_f, \theta_d) \\
& + \frac{1}{n'}\sum_{i=n+1}^{N}\mathcal{L}_d^{i}(\theta_f, \theta_d) ] ,
\end{split}
\end{align}
where $\theta_f, \theta_y, \theta_d$ are parameters of the joint feature extractor and two classifiers, and $\mathcal{L}_y, \mathcal{L}_d$ are the prediction and the domain loss functions. $n$ and $n'$ are the number of samples of the source and target domain data respectively.
We can optimize this loss function using stochastic gradient descent to get the parameters,
Using this DAT approach, we are able to minimize the divergence between the source and target feature distributions. Therefore, the learned embeddings are less dependent on the shift in language.

\subsection{Adversarial discriminative domain adaptation}

\begin{figure}[tbp]
    \centering
    \includegraphics[width=1.0\linewidth, height=4cm]{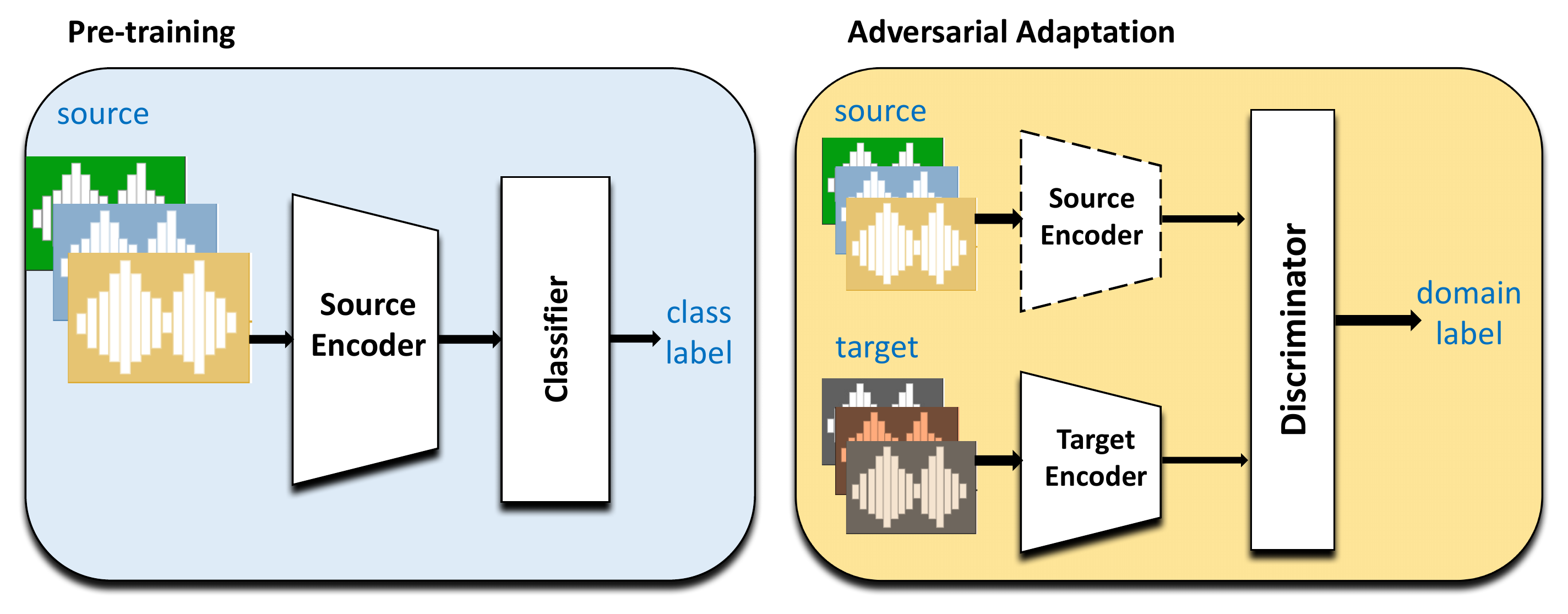}
    \caption{Overview of the proposed Adversarial Discriminative Domain Adaptation (ADDA) approach. Source DNN encoder is fixed during the adversarial adaptation.}
    \label{fig:adda}
\end{figure}

Different from the DAT method which applies a gradient reversal layer to confuse the domain classifier, we apply the Adversarial Discriminative Domain Adaptation (ADDA) approach to directly learn an asymmetric mapping, in which we modify the target model in order to match the source distribution. A summary of this entire training process is provided in Fig. \ref{fig:adda}.
Unlike the DAT method which uses a shared feature encoder, our proposed ADDA approach uses separate encoders for the source and target domain data. When there is a significant domain shift, the DAT method may not work well since it inherently assumes that source and target domain data has a similar class distribution.

We define input samples $\vct x \in \mat X$ with data labels $y \in \mat Y$, where $\mat X$ and $\mat Y$ are input space and output space, respectively. In our speaker verification experiments, $\vct x and y$ are x-vectors and speaker labels. The probabilistic distribution $\mathcal{D}(\vct x, y)$, however, might be different between training and evaluation dataset due to various domain mismatch such as language mismatch. We denote $\mathcal{S}(\vct x, y)$ and $\mathcal{T}(\vct x, y)$ as source domain and target domain distribution respectively. 
Our goal is to minimize the distance between the empirical source and target mapping distributions.
We firstly learn a source mapping $M_s$, along with a source classifier $\mathcal{C}$, and then learn to map the target domain encoder to the source domain.

We train the source classification model using a standard cross entropy loss defined below,
\begin{align}
\begin{split}
\min_{M_s, \mathcal{C}} \mathcal{L}_{cls} &(\mat X_s, Y_s) =  \\ 
& - \mathbb{E}_{(\vct x_s, y_s)\sim(\mat X_s, Y_s)} \sum_{k=1}^{K}\mathbbm{1}_{[k=y_s]}\log \mathcal{C}(M_s(\vct x_s)),
\end{split} 
\end{align}

In order to minimize the source and target representation distances,
we use a domain discriminator $\mathcal{D}$ to classify whether a data point is drawn from the source or the target domain. We optimize $\mathcal{D}$ using an adversarial loss $\mathcal{L}_{adv_D}(X_s, X_t, M_s, M_t)$, defined below: 
\begin{align}
\begin{split}
&\min_{D} \mathcal{L}_{adv_{D}}  (\mat X_s, \mat X_t, M_s, M_t) = \\ 
& - \mathbb{E}_{\vct x_s\sim \mat X_s} [\log \mathcal{D}(M_s(\vct x_s))] 
- \mathbb{E}_{\vct x_t\sim \mat X_t} [\log (1 - \mathcal{D}(M_t(\vct x_t)))], 
\end{split}
\end{align}

The DAT method uses a gradient reversal layer~\cite{ganin2016domain} to learn the mapping by maximizing the discriminator loss directly, where its adversarial loss $\mathcal{L}_{adv_M} = - \mathcal{L}_{adv_D}$. Different from DAT, in order to train the mapping, we use the loss function $\mathcal{L}_{adv_M}$ defined below. This objective has the same fixed-point properties as the minimax loss but provides stronger gradients to the target mapping.
\begin{align}
\min_{M_s, M_t} \mathcal{L}_{adv_{M}}(\mat X_s, \mat X_t, \mathcal{D}) &= - \mathbb{E}_{\vct x_t\sim \mat X_t} [\log \mathcal{D}(M_t(\vct x_t))].
\end{align}

We can optimize this objective function in two steps. First, we need to train a discriminative source classification model, we choose to use a three-layer Deep Neural Network (DNN) and the input features are x-vectors. We start optimizing classification loss $\mathcal{L}_{cls}$ over source domain mapping function $M_s$ and classifier $\mathcal{C}$ by training with the labeled source English data, $\mat X_s$ and $\mat Y_s$. Because we make $M_s$ fixed while learning $M_t$, we can then optimize $\mathcal{L}_{adv_D}$ and $\mathcal{L}_{adv_M}$ without revisiting the first objective term. 


Through this unsupervised adversarial discriminative domain adaptation approach, we can adapt the target encoder to the source domain. In the next section, we will present promising results on cross-lingual text-independent speaker verification tasks using ADDA.


\section{Experimental setup}
\label{sec:exp}

\subsection{English Corpora}
We use Speaker Recognition Evaluation (SRE) 04-08, Mixer 6, and Switchboard (SWBD) to train the x-vector model. SRE corpus is part of the Mixer 6 project, which was designed to support the development of robust speaker recognition technology by providing carefully collected speech across numerous microphones. Switchboard is a collection of about two-sided telephone conversations among thousands of speakers from all areas of the United States.

\subsection{Chinese Corpora}
AISHELL-1~\cite{bu2017aishell} is a subset of the AISHELL-ASR0009 corpus, which is a 500 hours multi-channel mandarin speech corpus designed for various speech/speaker processing tasks. Speech utterances are recorded at 44.1kHz via microphones, 16kHz via Android phones and 16kHz via iPhones.

There are 360 participants in the recording, and speakers' gender, accent, age, and birth-place are recorded as meta-data. 
About 80 percent of the speakers are from age 16 to 25. Most speakers come from the Northern area of China. The entire corpus includes training and test sets, without speaker overlap. Though the training data provides speaker labels, we do not use any speaker label information of the training data or include it in training our x-vector model. We only use it for unsupervised domain adaptation. We call it AISHELL unlabeled training set.

The training set contains 120,098 utterances from 340 speakers; 
Test set contains 7,176 utterances from 20 speakers. For each speaker, around 360 utterances (about 26 minutes of speech in total) are released. In order to test our proposed unsupervised ADDA approach, we don't use any speaker labels of the training data. We train our x-vector based speaker model on the SRE04-08, Mixer 6, and switchboard dataset, and evaluate on the Chinese AISHELL test 143520 trials. 

\subsection{Evaluation setup}
We use SRE04-08, Mixer6 and Switchboard data to train the TDNN based x-vector model. We follow the Kaldi SRE16 recipe to augment the training data by adding noises and reverberations. We use an energy based VAD and the raw feature to train the model are 23-dimensional MFCCs. 
Having established the x-vector system using English data, we now try to address the challenge of evaluation enrollment and test speakers for a mismatched language, Chinese. To accomplish this,
A set of unlabeled data for the new language is needed. We use the target domain AISHELL unlabeled training data.
We extract x-vectors on source domain SRE and SWBD data and target domain AISHELL unlabeled data to train the adaptation network.

We train the Adversarial Domain Adaptation Network (ADAN) in two steps. First, we train a DNN encoder and classifier on SRE and SWBD x-vectors. Next, we use the pre-trained source model as an initialization for the target DNN encoder and perform adversarial adaptation to learn a target domain mapping on the AISEHLL unlabeled x-vectors.

During testing, we use AISHELL evaluation set enrollment x-vectors and test x-vectors as the input to the ADDA, and extract the new vectors $\hat{x}_e$, $\hat{x}_t$ using the trained target encoder of ADDA.  Adapted embeddings $\hat{x}_e$, $\hat{x}_t$ are therefore expected to be domain-invariant and speaker discriminative representations which stay in the same subspace. We apply mean and length normalization on the adapted embeddings. For the back-end, we train a Probabilistic Linear Discriminant Analysis (PLDA) model on combined SRE clean and noise augmented data, and compute log-likelihood ratio scores of enrollment and test trials. We also perform unsupervised PLDA adaptation using Kaldi to utilize the AISHELL unlabeled data.

\subsection{Model configuration}

For this experiment, our base architecture is a three-layer Deep Neural Network which is fine-tuned on the source domain for 100 epochs using a batch size of 128. When training ADDA, the adversarial discriminator consists of three additional fully connected layers: 2 hidden layers and an adversarial discriminator output. With the exception of the output, these additionally fully connected layers use a ReLU activation function. ADDA target encoder training then proceeds for another 100 epochs with a batch size of 128. For the DAT training, the shared feature encoder is a three-layer DNN. We use an Adam optimizer with a learning rate $10^{-4}$. The speaker classifier and the language classifier are two-layer DNNs. To confuse the language domain classifier, the language classifier has a gradient reversal layer. We use a multi-task loss with equal weights to combine the two cross entropy losses.



\section{Results and Discussions}
\label{sec:result}

\subsection{Results}
In this section, we show experimental results using x-vector, x-vector with DAT and x-vector with ADDA training with and without PLDA adaptations in \cref{tab:eer}.  We use Linear Discriminant Analysis (LDA) to reduce all three embeddings to 256 dimension for comparison. Also, we concatenate the DAT embedding with the x-vector since we find it always performs better than a single DAT embedding.
From \cref{tab:eer}, we observe that our proposed method, ADDA, greatly improves Equal Error Rate (EER) on AISHELL test trials. After ADDA adaptation, EER of the x-vector system decreases from 9.331\% to 7.645\%, relatively 18.07\%. The ADDA approach also achieves relatively 12.54\% improvement compared with the concatenated x-vector and DAT embedding. The major reason that ADDA works better might be that it uses an adversarial discriminator to adapt the target encoder to the source domain. Also, by initializing the target representation space with the pre-trained source model, we can effectively learn the asymmetric mapping function.

\begin{figure}[htbp]
    \centering
    \includegraphics[width=0.85\linewidth, height=4.6cm]{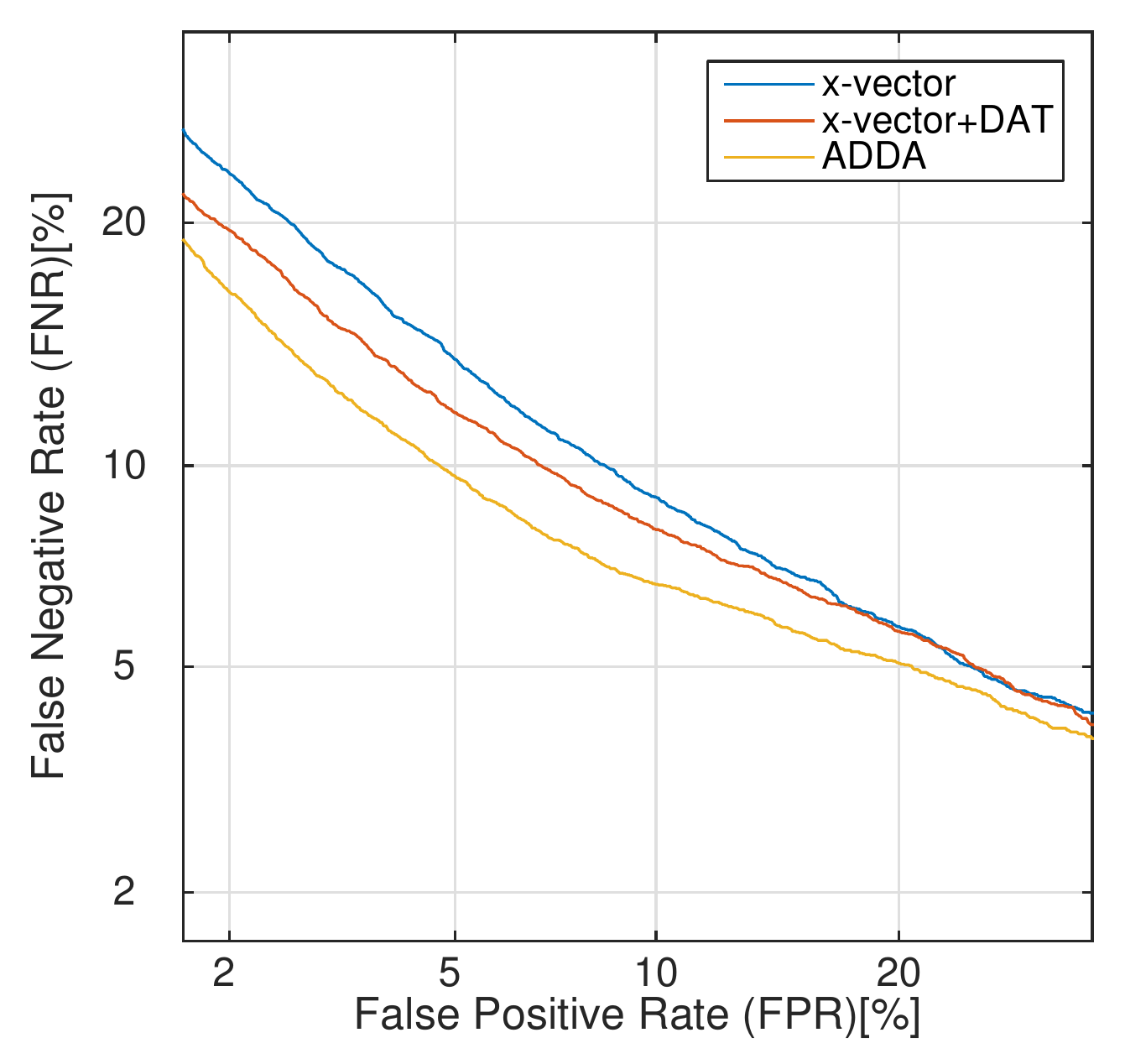}
    \caption{DET curve results with different speaker representations.}
    \label{fig:det}
\end{figure}

\cref{fig:det} shows the  Detection Error Trade-off (DET) curve of our speaker recognition system at three different settings without PLDA adaptation. From the figure, we see after DAT or ADDA adaptation, the overall speaker verification system performance improves significantly compared with the x-vector system. Further, both False Positive Rate (FPR) and False Negative Rate (FNR) of the ADDA embedding system reduce by a large margin compared with the x-vector+DAT embedding system. It indicates that ADDA embedding has more invariance to language shift.

\vspace{-2ex}

\begin{table}[htbp]
  \centering
    \begin{tabular}{lrrr}
        \toprule

          & \multicolumn{1}{l}{EER(\%)} & \multicolumn{1}{l}{MinDCF} &  \\
    \midrule

    x-vector & 9.331 & 0.7755 &  \\
    x-vector + DAT & 8.741 & 0.7475 &  \\
    ADDA embedding &  7.645 &  0.7257 &  \\
    \midrule
    x-vector + PLDA adaptation & 9.162 &  0.7095 &  \\
    x-vector + DAT + PLDA adaptation  & 7.799 & 0.6989 &  \\
    ADDA embedding + PLDA adaptation &  7.504   & 0.7062 &  \\

    \bottomrule

    \end{tabular}%
      \caption{Speaker verification results using different models with a PLDA back-end.}

  \label{tab:eer}%
\end{table}%
\vspace{-3ex}

\subsection{Visualization of speaker embeddings}

\begin{figure}[tbp]

\begin{minipage}[b]{.48\linewidth}
  \centering
  \centerline{\includegraphics[width=4.5cm]{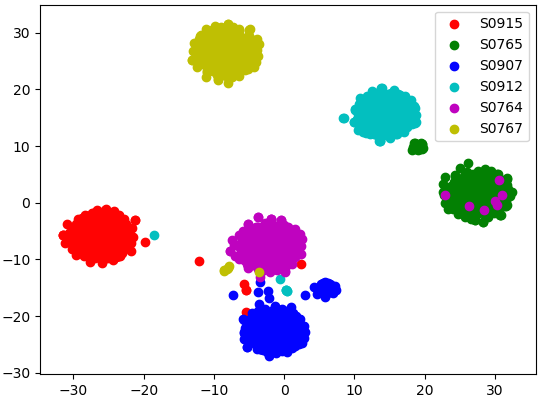}}
  \centerline{(a) x-vector}\medskip
\end{minipage}
\hfill
\begin{minipage}[b]{0.48\linewidth}
  \centering
  \centerline{\includegraphics[width=4.5cm]{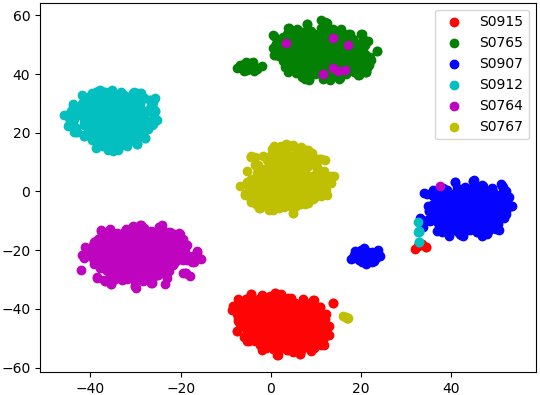}}
  \centerline{(b) ADDA embedding}\medskip
\end{minipage}
\caption{Visualizations of x-vector and ADDA speaker embeddings using t-SNE}
\label{fig:tsne}
\end{figure}

To investigate the effect ADDA has on speaker verification, we further assess the quality of the learned speaker features, using t-SNE~\cite{maaten2008visualizing}, we plot embeddings after LDA from same K speakers of the AISHELL test set.  The results are presented in \cref{fig:tsne}. \cref{fig:tsne} (a) is the visualization of x-vectors, and \cref{fig:tsne} (b) is the visualization of ADDA embedding. It can be seen that the ADDA embeddings have more discriminative ability to separate different speakers. However, for x-vectors, we observe that some utterances from different speakers are grouped together and not well separated in the embedding space. Also, for speaker ``0764'', it is difficult to separate it from speaker ``0765'' using both methods. It is probably because these two speakers have very similar speaker information.

\subsection{Clustering analysis}
In order to quantitatively analyze the quality of adapted speaker representations, we also perform clustering on the adapted embeddings. Since t-SNE cannot maintain distance information, which is necessary to apply most clustering algorithms, we perform K-means clustering after LDA transformed x-vectors and ADDA embeddings. Given the knowledge of the ground truth speaker labels, we compute the Normalized Mutual Information (NMI)~\cite{vinh2010information} of the K-means clustering assignment. NMI is a metric that measures the agreement of the ground truth labels and the clustering results. The NMI score of x-vectors is 0.787, and the NMI score of ADDA embeddings is 0.802, relatively 1.9\% higher. This result is consistent with the visualization using t-SNE. Therefore, we can conclude that with the ADDA adaptation, we can learn more speaker discriminative and language independent speaker embeddings.

\section{Conclusions and Future Work}
\label{sec:conclusion}

We presented a discriminative adversarial unsupervised adaptation method in this paper.
  By exploiting how to alleviate the domain mismatch problem in an English-Chinese cross-lingual speaker verification task, we showed that our proposed unsupervised ADDA approach can perform well on speaker classification for the target domain data. Additional data analysis indicated that the representations learned via ADDA can be well separated and are less dependent with respect to the shift in language. 

  In the future, we would like to investigate the influence of phonetic content on cross-lingual text-independent speaker verification. We intend to use a phoneme decoder to analyze the linguistic factor of speaker models.

\vfill\pagebreak

\newpage

\bibliographystyle{IEEEbib}
\bibliography{icassp18_adv.bib}

\end{document}